\documentclass[11pt, twoside, usenatbib, usepslatex]{article}

\usepackage{asp2006}
\usepackage{epsf}
\usepackage{psfig}
\usepackage{graphicx}
\usepackage{natbib}
\usepackage{subfigure}

\begin{document}

\title{The inhomogeneous response across the solar disc of unresolved Doppler velocity observations}

\author{A.-M. Broomhall,$^1$ W.~J. Chaplin,$^1$
Y. Elsworth$^1$ and R. New$^2$}

\affil{$^1$School of Physics and
Astronomy, University of Birmingham, Edgbaston, Birmingham B15 2TT\\
$^2$Faculty of Arts, Computing, Engineering and Sciences, Sheffield Hallam University, Sheffield S1 1WB}    %%% Fill in author affiliations

\begin{abstract}
Unresolved Doppler velocity measurements are not homogenous across
the solar disc \citep{Brookes1978}. We consider one cause of the
inhomogeneity that originates from the BiSON instrumentation itself:
the intensity of light observed from a region on the solar disc is
dependent on the distance between that region on the image of the
solar disc formed in the instrument and the detector. The
non-uniform weighting affects the realization of the solar noise and
the amplitudes of the solar oscillations observed by a detector. An
`offset velocity', which varies with time, is observed in BiSON data
and has consequences for the long-term stability of observations. We
have attempted to model, in terms of the inhomogeneous weighting,
the average observed offset velocity.
\end{abstract}
\section{Introduction}
Birmingham Solar Oscillations Network (BiSON) instruments use
resonant scattering spectrometers to make unresolved Doppler
velocity observations of the Sun. This involves determining the
difference between the intensity of light observed on the red and
blue wings of a solar Fraunhofer line. Sun-as-a-star Doppler
velocity measurements are not homogenous across the solar disc and
so the observed data do not represent a uniform average over the
entire surface. The inhomogeneity is influenced by the solar
rotation and limb darkening \citep{Brookes1978}. We have considered
a further instrumental effect that occurs because the image is
viewed through a vapour: the intensity of light observed from a
particular region on the solar disc is dependent on the position of
the detector with respect to the image of the Sun seen by the
instrument. The majority of BiSON instruments have two detectors,
positioned on opposite sides of the observed solar image. The
observations made by each detector are weighted towards differing
regions of the solar disc, which affects the observed mode
amplitudes and granulation noise.

The layout of the rest of this paper is as follows. In Section 2, we
model the bias across the solar disc that is seen by BiSON
instruments by accounting for the position of the detector with
respect to the observed image of the Sun. In Section 3 we measure
and attempt to model an `offset velocity' that is present in the
BiSON data and the results are summarised in Section 4.
\section{Modelling the bias across the disc}\label{section[model bias]}
\begin{figure}
  \centering
  \includegraphics[trim = 0mm 15mm 0mm 0mm, width=0.44\textwidth, clip]{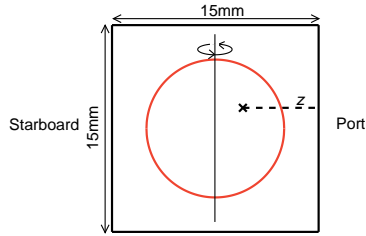}\\
  \vspace{-0.2cm}
  \caption{A schematic of a vapour cell with the port and starboard detectors on either side. Also depicted
  are the axis and direction of rotation of the observed solar image (red circle). We take the optical path
  of light to the detector to be horizontal, as shown by the dashed line, and $z$ is
  the distance travelled through the vapour by light.}\label{figure[vapour_cell]}
\end{figure}

The port and starboard detectors of a BiSON instrument are
positioned on either side of a potassium vapour cell (see Figure
\ref{figure[vapour_cell]}). The intensity observed by the detector
from a given region on the Sun, $I$, is determined by the optical
depth of the vapour in the cell, $\tau$, and is given by $I =
I_0\textrm{e}^{-\tau}$ , where $I_0$ is the intensity of light
received from the Sun. The optical depth of a vapour can be
described as $\tau = Kz$, where $K$ is the extinction coefficient
and $z$ is the optical path. Here, as a first-order approximation,
we have only considered the perpendicular distance between the
region of the image and the inside wall of the vapour cell (as shown
in Figure \ref{figure[vapour_cell]}). We assume that $K$
is constant throughout the vapour cell. We now use this model to
determine the observed bias across the solar disc.

\begin{figure*}
  \centering
  \subfigure[Blue wing, starboard
  detector]{\includegraphics[width=0.37\textwidth, clip, trim=-27mm 0mm -27mm 0mm]{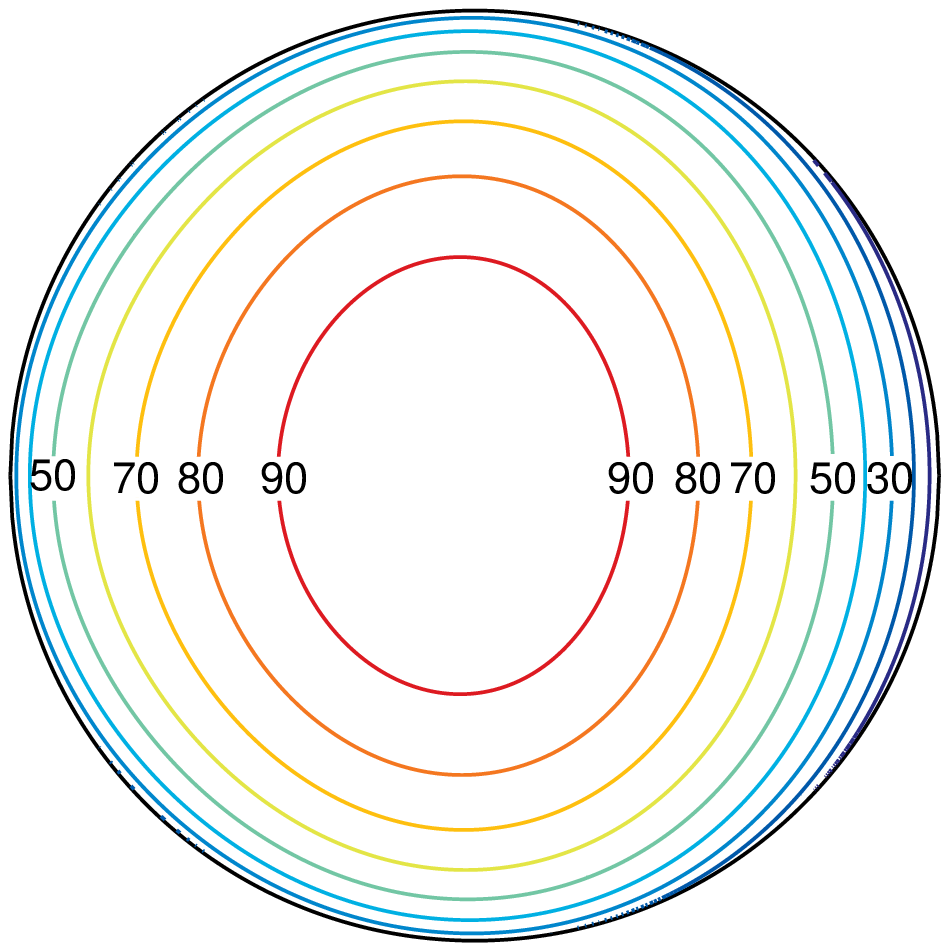}}
  \hspace{1cm}
  \subfigure[Red wing, starboard detector]{\includegraphics[width=0.37\textwidth, clip, trim=-27mm 0mm -27mm 0mm]{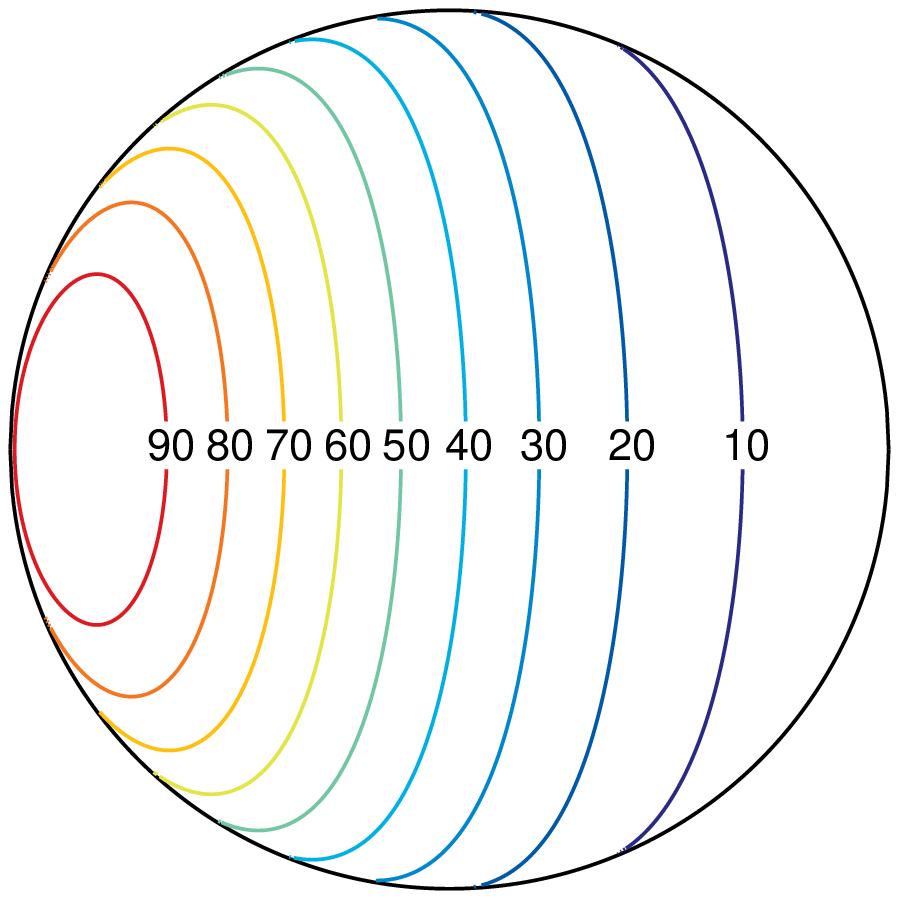}}\\
  \subfigure[Blue wing, port
  detector]{\includegraphics[width=0.37\textwidth, clip, trim=-27mm 0mm -27mm 0mm]{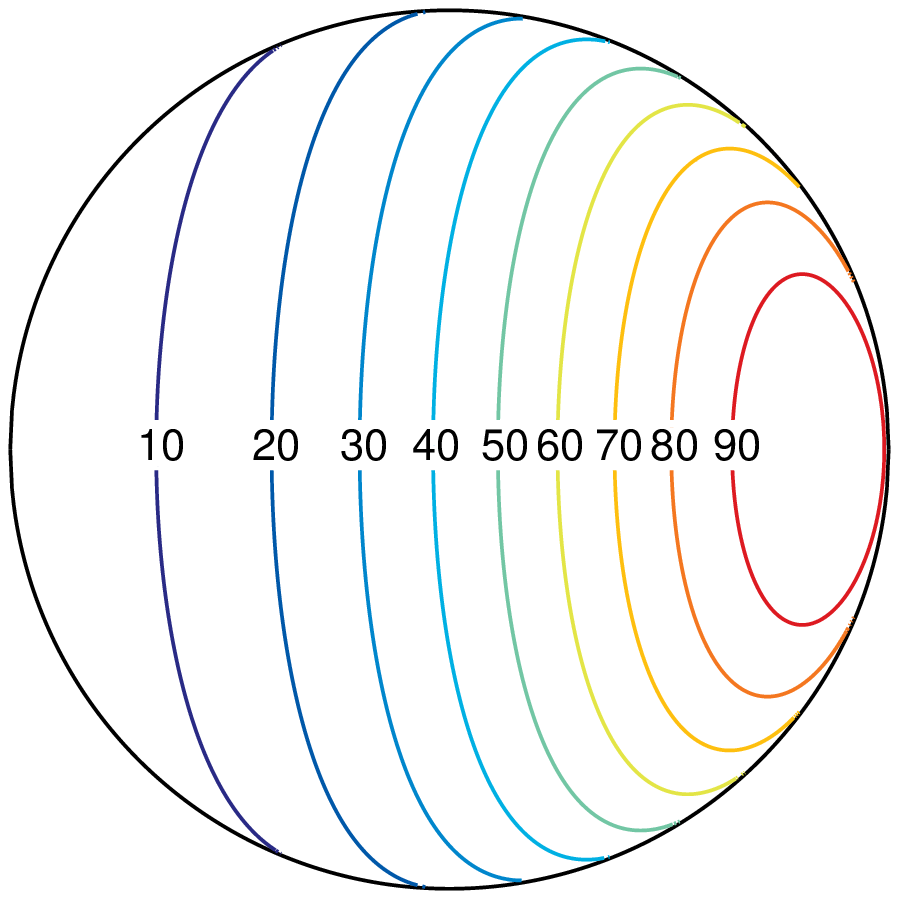}}
  \hspace{1cm}
  \subfigure[Red wing, port detector]{\includegraphics[width=0.37\textwidth, clip, trim=-27mm 0mm -27mm 0mm]{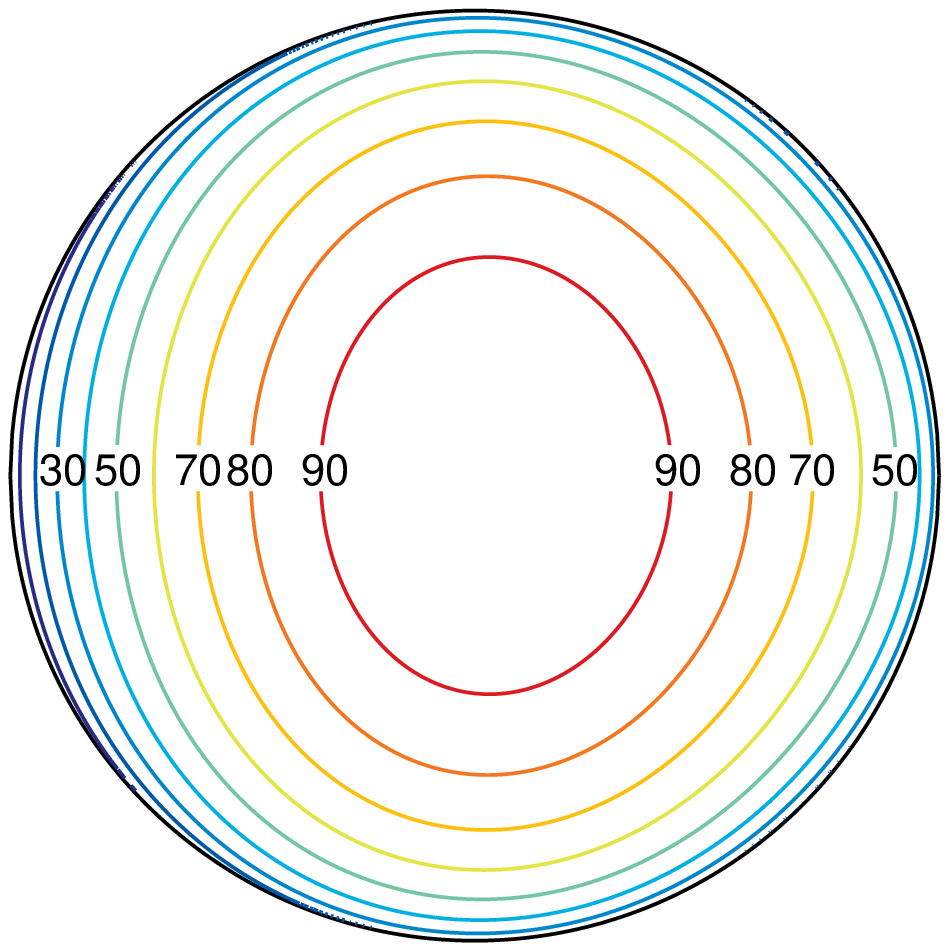}}\\
  \caption{Contour maps showing the non-uniform weighting of the solar disc.
  The intensities have been scaled so that the region which contributes
  the most has been given a weight of 100, while the regions that contribute
  the least have been given a weight of 0 and contours in the range 10 to
  90 have been plotted. In the contour maps the left limb is approaching
  an observer and the right limb is receding.}\label{figure[all contour vlos0]}
\end{figure*}

We have combined the effects of the solar rotation and limb
darkening with the weighting caused by the position of the detector
to produce contour maps of the weighting of the solar disc (see
Figure \ref{figure[all contour vlos0]}). The instrumental effect
means that each combination of wing and detector is weighted towards
differing regions of the solar disc and so the sensitivity to the
observed modes of each combination is different. Each detector will
observe at a different height in the solar atmosphere, affecting the
amplitudes of the modes and the realisation of the granulation noise
observed by the detector. The weighting pattern is dependent on the
line-of-sight velocity between the Sun and the observing instrument
(`the station velocity'), which was set to $0\,\rm m\rm s^{-1}$ for
the shown contour maps. As the station velocity increases the
weightings are shifted towards the approaching limb of the Sun.

We now use this image of the weighting of the Sun to try and model
an offset velocity that is observed in BiSON data.

%%%%%%%%%%%%%%%%%%%%%%%%%%%%%%%%%%%%%%%%%%%%%%%%%%%%%%%%%%%%
%%%%%%%%%%%%%%%%%%%%%%%%%%%%%%%%%%%%%%%%%%%%%%%%%%%%%%%%%%%%%%%%%%%%%%%%%%%%%%%%%%%%%%%%%%
\section{Velocity offset observed in BiSON data}\label{section[velocity offset]}
\begin{figure}
  \centering
  \subfigure{\includegraphics[height=3.7cm,
  clip]{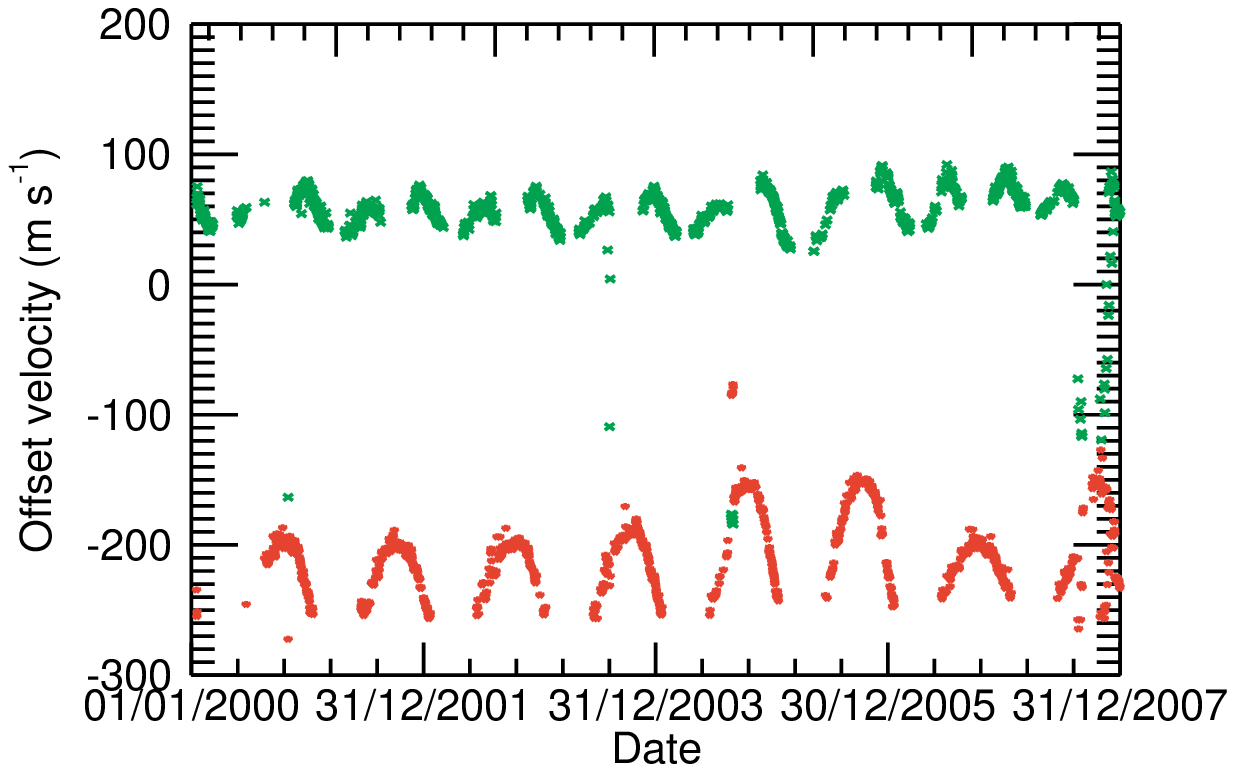}}\hspace{0.2cm}
  \subfigure{\includegraphics[height=3.7cm, clip]{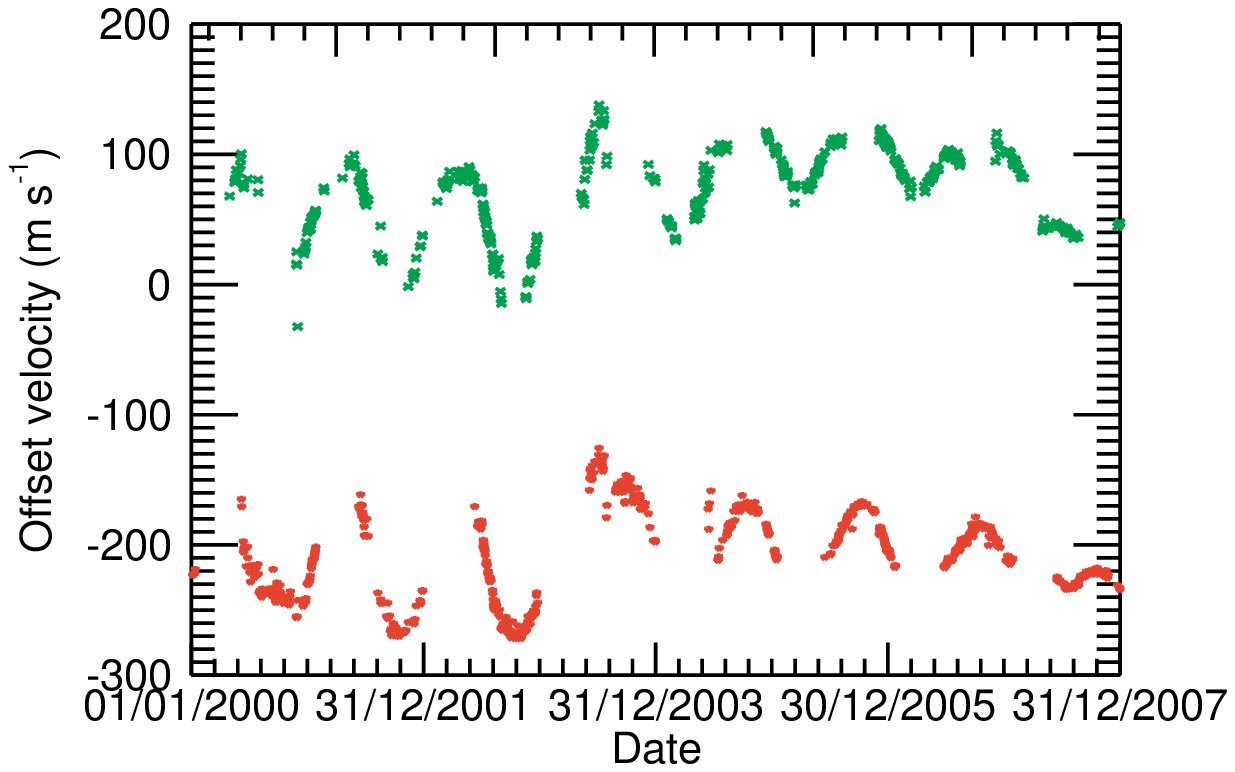}}\\
  \vspace{-0.2cm}
  \caption{The offset velocity observed in Las Campanas (left panel)
  Narrabri (right panel) data. The red triangles represent
  the port detector offsets, while the green crosses show
  the starboard detector offsets.}\label{figure[obs vel offset]}
\end{figure}

The observed velocity can be extracted from the raw intensity
measurements by determining the ratio, $R$, which is given by
$R=(I_b-I_r)/(I_b+I_r)$, where $I_b$ and $I_r$ are the strengths of
the resonantly scattered signal on the blue and red wings of the
solar potassium absorption line respectively. The ratio varies with
station velocity but a given value of the ratio should correspond to
the same velocity from day to day. We have determined, for many
different days, the value of the observed velocity that corresponds
to a ratio of $R=0.15$ and we call this determined velocity the
offset velocity (see Broomhall at al., in preparation, for further
details). The observed offset velocity is different for the port and
starboard observations and varies systematically with time (see
Figure \ref{figure[obs vel offset]}). The observed offset velocity
is analogous for the different BiSON instruments.

%%%%%%%%%%%%%%%%%%%%%%%%%%%%%%%%%%%%%%%%%%%%%%%%%%%%%%%%%%%%%%%%%%%
%%%%%%%%%%%%%%%%%%%%%%%%%%%%%%%%%%%%%%%%%%%%%%%%%%%%%%%%%%%%%%%%%%%

\begin{figure}
  \centering
  \includegraphics[height=3.7cm, clip]{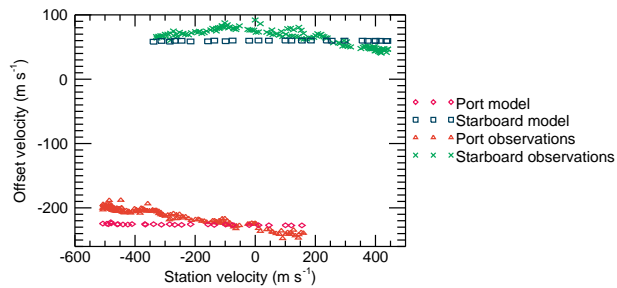}\\
  \vspace{-0.2cm}
  \caption{A comparison between the model offset velocity and the
  offset velocity observed in 2006 by the Las Campanas instrument.
  The results are plotted as a function of the station velocity.}\label{figure[model offset]}
\end{figure}

We have used our model of the observed inhomogeneity to determine
whether this effect can be recreated in artificial data and the
results are shown in Figure \ref{figure[model offset]}. In the model
we took the optical depth at 15mm to be $\tau=1.1$. The average
observed and model offset velocities are in reasonably good
agreement, however, the observed offset velocity shows significantly
more variation than the model offset. In the model we have varied
the orientation of the rotation axis and the size of the observed
image with time, which introduces some variation into the offset
velocity. However, the range is still at least a factor of 10 too
small and to date we have no satisfactory explanation for this.

\section{Discussion}\label{section[discussion]}
We have shown that the observed weighting across the solar disc is
different for the red and blue wing observations and the port and
starboard observations. The addition of the instrumental effect
allows us to model the difference that exists between the port and
starboard observations. The observations are biased towards
different portions of the solar Fraunhofer line and so each
combination of wing and detector observes at a different height in
the solar atmosphere. Therefore, the realizations of the solar noise
observed by each combination of wing and detector differs slightly,
as will the observed amplitudes of the solar oscillations. Since the
position of the detector has a significant influence on the
instrument's sensitivity across the solar disc it also affects the
visibility of the mode \citep{Christensen1989}.

An offset velocity is observed, in BiSON data, to vary by of the
order of $50\,\rm m\,s^{-1}$ over the course of a year. The observed
offset velocity introduces low-frequency noise into BiSON data and
affects the long-term stability of the observations. Although the
model of the observed weighting produced is able to make reasonably
good predictions of the average offset velocities it is unable to
explain the large observed variation with time.

\acknowledgements We are grateful to R. Simoniello for providing
Themis data. We thank the members of the BiSON team, both past and
present, and all those involved in the data collection process. The
authors acknowledge the financial support of the Science and
Technology Facilities Council (STFC).

\bibliographystyle{mn2e}
\bibliography{weighting_ref}

\begin{thebibliography}{}

\bibitem[\protect\citeauthoryear{Brookes, Isaak \& van~der Raay}{Brookes
  et~al.}{1978}]{Brookes1978}
Brookes J.,  Isaak G.,    van~der Raay H.,  1978, MNRAS, 185, 19

\bibitem[\protect\citeauthoryear{{Christensen-Dalsgaard}}{{Christensen-Dalsgaa%
rd}}{1989}]{Christensen1989}
{Christensen-Dalsgaard} J.,  1989, MNRAS, 239, 977

\end{thebibliography}

\end{document}